Comments on the Paper

# 'A new basic 1-dimensional 1-layer model obtains excellent agreement with the observed Earth temperature'

by Rainer Link and Horst-Joachim Lüdecke

## Gerhard Kramm[1] and Ralph Dlugi[2]

[1]University of Alaska Fairbanks, Geophysical Institute
903 Koyukuk Drive, P.O. Box 757320, Fairbanks, AK 99775-7320, USA

[2]Arbeitsgruppe Atmosphärische Prozesse (AGAP),
Gernotstraße, D-80804 Munich, Germany

In our comments on the paper of Link and Lüdecke we document that these authors used rather improper quotations of our paper. They also argued on the basis of false claims regarding our mathematical and physical description of both the global energy balance model of Schneider and Mass and the Dines-type two-layer energy balance model for the Earth-atmosphere system. They completely disregarded the respective literature. They claimed excellent agreement between their predicted Earth's surface temperature and the observed one even though they only reproduced the temperature by an inverse application of the power law of Stefan and Boltzmann. They also claimed that their result for the increase in the Earth's surface temperature of $\Delta T \approx 1.1 \, \text{K}$ due to the doubling of the atmospheric $CO_2$ concentration is in good agreement with the IPCC value if no feedback is considered. However, beside the fact that this result is not based on model predictions, it also disagrees with the definition of the anthropogenic radiative forcing as can be found not only in the respective literature, but also in various IPCC reports.

Keywords: Global energy balance models; scientific standards; Earth-atmosphere system.

PACS Nos.: 51.30.+I, 42.68.Ay, 92.60.Fm, 92.60.Kc, 92.60.Ry, 92.60.Ta

## 1. Introduction

With interest we read the paper of Link and Lüdecke[1] (hereafter abbreviated by LL). Unfortunately, their paper contains rather improper quotations of our paper[2] (hereafter abbreviated by KD). LL also argued on the basis of false claims regarding our mathematical and physical description of both the global energy balance model of Schneider and Mass[3] and the Dines-type two-layer energy balance model for the Earth-atmosphere system.[4] LL even disregarded the model descriptions of those authors. Furthermore, the paper of LL is based on various fundamental physical and mathematical mistakes. Their claim that their new basic 1-dimensional 1-layer model predicts the Earth's surface temperature which is in excellent agreement with the observed one is rather inadequate because they only reproduced the temperature input by an inverse application of the power law of Stefan[5] and Boltzmann.[6] Note that their model is neither a one-dimensional model because no spatial coordinate occurs in the governing equations nor a one-layer model because it consists of two layers: (a) one layer beneath the Earth's surface representing the upper layer of an aqua-planet, and (b) one layer above the Earth's surface representing the atmosphere. Moreover, the model of LL exhibits distinct inconsistencies that cause results which completely disagree with those listed in the respective literature. LL, for instance, ignored that infrared radiation is an example of isotropic radiation, a fundamental prerequisite in deriving the power law of Stefan and Boltzmann. This power law, however, is used to relate various radiative fluxes emitted by the Earth's surface and the atmosphere to the corresponding temperatures.

Therefore, it is indispensable (a) to correct their improper quotations, (b) to rebuff their false claims, and (c) to document their fundamental physical and mathematical mistakes and their distinct model inconsistencies for the purpose of clarification.

## 2. False claims

### 2.1 The energy balance model of Schneider and Mass

The global energy balance model of Schneider and Mass[3] expressed by Eq. (2.1) of KD reads (see also Eq. (1) of LL):

$$R \frac{dT_s}{dt} = \left(1 - \alpha_E\right) \frac{S}{4} + \varepsilon_E \, F_{IR\downarrow} - F_{IR\uparrow}\left(T_s\right) \quad . \tag{2.1}$$

Here, $R$ is the planetary inertia coefficient, i.e., the heat capacity (of water) times the thickness of the water layer of an aqua-planet as considered by Schneider and Mass[3], $T_s$ is the surface temperature of this water layer, $t$ is time, $S$ is the solar constant, $\alpha_E \leq 1$ and $\varepsilon_E \leq 1$ are the planetary integral albedo and the planetary integral emissivity of the Earth, respectively. Furthermore, the quantity $F_{IR\downarrow}$ is the flux density (hereafter simply called a flux) of down-welling infrared (IR) radiation, and $F_{IR\uparrow}\left(T_s\right)$ is the IR radiation emitted by the Earth's surface. On page 450 of their article, LL claimed:

> »They finally improved the model of Schneider and Mass by including the sensible and latent heat and the absorption of solar radiation by the atmosphere. However, also this corrected model shows substantial drawbacks. Kramm and Dlugi used only one temperature $T_a$ and one integral emissivity $\varepsilon_a$ for the lower and upper part of the atmospheric layer (Eq. A19, A20 in Ref. 3). This is not feasible because of the big difference between the long-wave radiation from the top of the atmosphere and the down-welling infrared radiation to the Earth at the lower atmosphere.«

These claims of LL are false. The global energy balance model of Schneider and Mass[3] does not contain a temperature, $T_a$, and/or an integral emissivity, $\varepsilon_a$, of the atmosphere. This is even true in case of the version extended by us.[2]

Schneider and Mass[3] did not use the power law of Stefan and Boltzmann, neither for $F_{IR\downarrow}$ nor for $F_{IR\uparrow}(T_s)$. Instead, they used Budyko's empirical formula given by[7, 8]

$$\Delta F_{IR} = F_{IR\uparrow}(T_s) - \varepsilon_E \, F_{IR\downarrow} = a + b\,(T_s - T_r) - \left\{ a_1 + b_1\,(T_s - T_r) \right\} \, n \quad , \tag{2.2}$$

where the effect owing to clouds expressed by the normalized cloud cover $n$ was ignored. Here, $a = 226.0 \, \text{W m}^{-2}$, $b = 2.26 \, \text{W m}^{-2} \, \text{K}^{-1}$, $a_1 = 48.4 \, \text{W m}^{-2}$ and $b_1 = 1.61 \, \text{W m}^{-2} \, \text{K}^{-1}$ are empirical constants, and $T_r = 273.15 \, \text{K}$ is a reference temperature. Note that the term $\Delta F_{IR} = F_{IR\uparrow}(T_s) - \varepsilon_E \, F_{IR\downarrow}$ is called the net radiation flux in the infrared range. Combining Eqs. (2.1) and (2.2) yields for clear-sky conditions ($n = 0$):[2]

$$R \, \frac{dT_s}{dt} = Q - \lambda \, T_s \tag{2.3}$$

with the thermal forcing

$$Q = (1 - \alpha_E) \frac{S}{4} - a + b \, T_r \tag{2.4}$$

and $\lambda = b$ called the feedback parameter. Since Schneider and Mass[3] expressed the solar constant, $S$, occurring in Eq. (2.4), as a function of time due to the time-dependent influence of solar activity (characterized by sunspot numbers) and atmospheric dust (provided, for instance, by volcano eruptions), they presented results from numerical solutions of equation set (2.3) and (2.4).

If we, however, assume that $\alpha_E$, $\varepsilon_E$, and $S$, and, hence, $Q$ are independent of time as assumed by various authors,[9-11] the exact solution of Eq. (2.3) is given by[2]

$$T_s(t) = T_{s0} \exp\left(-\frac{\lambda}{R} \, t\right) + \frac{Q}{\lambda} \left(1 - \exp\left(-\frac{\lambda}{R} \, t\right)\right) \quad , \tag{2.5}$$

where $T_{s0} = T_s(t = 0)$ is the initial surface temperature. If time tends to infinity, we will obtain: $T_s(\infty) = T_s(t \to \infty) = Q/\lambda$ also called the radiative equilibrium temperature[9] or the fixed point temperature that is based on the condition: $dT_s/dt = 0$.

As reflected by Figure 1, such a radiative equilibrium does not exist at the Earth's surface. Consequently, we inserted the globally averaged fluxes of sensible heat, $H$, and latent heat, $E = L_v(T_s) \, W$, at the surface of the water layer into Eq. (1), where $L_v(T_s)$ is the specific heat of phase transition (e.g., vaporization, sublimation), considered as dependent on the surface temperature, $T_s$, and $W$ is the water vapor flux.[2] In addition, we considered the globally averaged atmospheric absorption of solar radiation expressed by the, $A_a S/4$, where $A_a$ is the integral absorptivity with respect to the range of solar radiation. Thus, we obtained (see also Eqs. (2.16) and (A17) in Ref. 2):

$$R \frac{dT_s}{dt} = \left(1 - \alpha_E - A_a\right) \frac{S}{4} - H - E - \Delta F_{IR} \quad . \tag{2.6}$$

Using Eq. (2.2) for clear-sky conditions leads to Eq. (2.3) again, but the forcing term (2.4) becomes

$$Q = \left(1 - \alpha_E - A_a\right) \frac{S}{4} - H - E - a + b\, T_r \quad . \tag{2.7}$$

Obviously, neither the global energy balance model of Schneider and Mass[3] nor the extended form (Eqs. (2.6) and (2.7)) exhibits any temperature, $T_a$, and/or any integral emissivity, $\varepsilon_a$, of the atmosphere (for more details, see section 2 of Ref. 2). Thus, LL tried to falsify both the paper of Schneider and Mass[3] and that of KD.

In contradiction to the claim of LL, Eqs. (A18) and (A19) are related to the Dines-type two-layer energy balance model described in section 6 of KD. The attempt of LL to relate these two equations to the global energy balance model of Schneider and Mass[3] is, therefore, inconceivable.

In section 5 of our paper[2] we discussed Ångström-type and Brunt-type formulae for the down-welling IR radiation that are based on long-term observations. These formulae are given by[2]

$$F_{IR\downarrow} = \sigma\, T_L^4 \begin{cases} \left(a_3 - b_3\, 10^{-\gamma e_L}\right) & \text{if Ångström} - \text{type formulae}^{13,\,14} \\[2mm] \left(a_4 + b_4\, e_L^{\frac{1}{2}}\right) & \text{if Brunt} - \text{type formulae}^{14,\,15} \end{cases} \quad . \tag{2.8}$$

Here, $T_L$ is the air temperature close to the surface (at the height of 1.5 to 2 m at which the measurements are performed using various thermometers being located in so-called weather huts), $e_L$ is the water vapor pressure at the same height either in mm Hg (Ångström) or in hPa (Brunt), and $a_3$, $b_3$, $\gamma$, $a_4$, and $b_4$ are empirical constants (listed, for instance, in Table 2 of Ref. 2). As shown by KD, these empirical formulae may be multiplied by a weighting function for both the cloud cover and the cloud type given by[16]

$$f(n) = 1 + K\, n^{2.5} \quad , \tag{2.9}$$

where $K$ is a weighting factor that depends on the cloud type. It is ranging from 0.04 for cirrus clouds to 0.24 for stratus clouds; an average value is given by $K \approx 0.22$ (see Ref. 16).

Since the temperatures $T_L$ and $T_s$ may notably differ from each other, we may express the former by the latter according to

$$T_L = T_s - \Delta T \quad , \tag{2.10}$$

where the deviation $\Delta T$ may be either positive or negative. Thus, the term $\Delta F_{IR} = F_{IR\uparrow}\left(T_s\right) - \varepsilon_E \, F_{IR\downarrow}$ in Eq. (2.6) results in

$$\Delta F_{IR} = \sigma \, T_s^4 \begin{cases} \varepsilon_E \left\{ 1 - \left(1 - 4 \, \delta T\right)\left(a_3 - b_3 \, 10^{-\gamma \, e_L}\right)\left(1 + K \, n^{2.5}\right)\right\} & \text{if Refs. 13, 14, and 16} \\ \varepsilon_E \left\{ 1 - \left(1 - 4 \, \delta T\right)\left(a_4 + b_4 \, e_L^{\frac{1}{2}}\right)\left(1 + K \, n^{2.5}\right)\right\} & \text{if Refs. 14 to 16} \end{cases} , \tag{2.11}$$

where $\delta T = \Delta T / T_s$ is considered as a relative deviation. Thus, we only used (near-)surface values for the temperature and the humidity. Therefore, an atmospheric temperature, $T_a$, was not considered because a mean atmospheric temperature is rather inappropriate in case of the Ångström-type and Brunt-type formulae.

It should be noticed that Link and Lüdecke already wrote a comment on the paper of KD in December 2010.[17] Figure 1 of our reply to this comment[18] exhibits the page 743 of the paper of Schneider and Mass[3] in which the description of their global energy balance model is given. This model description documents that no atmospheric temperature occurs in the model of Schneider and Mass[3]. Obviously, LL knew the truth, at least, since the beginning of January 2011, i.e., three months before they submitted the manuscript to the International Journal of Modern Physics C.

### 2.2 The direct emission of long-wave radiation by the Earth's surface into the space
On the same page of their article, LL continued their false claims by:[1]

>Furthermore, the direct emission of the long wave radiation from Earth to space was wrongly taken into account by a term

$$-\left(1 - \varepsilon_a\right) \varepsilon_E \, \sigma \, T_s^4$$

(pp. 152 and Eq. A18 in Ref. 3) that disappears for $\varepsilon_a = 1$. This was the reason why Kramm and Dlugi obtained for the mean Earth temperature $T_E = 268$ K (Figure 15 in Ref. 3) that disagrees with the observed value $T_E = 288$ K. Kramm and Dlugi calculated the mean temperature of the atmosphere as $T_A = 255$ K. In their paper this value is independent of the integral absorptivity of the atmosphere with respect to the range of solar radiation ($A_a$ in Fig. 15 of Ref. 3), which seems rather odd.«

First at all, the term $-\left(1-\varepsilon_a\right)\varepsilon_E\ \sigma\ T_s^{\ 4}$ does not occur in the global energy balance model of Schneider and Mass[3], neither in the original version (see Eqs. (2.3) and (2.4)) nor in the extended one[2] (see Eqs. (2.6) and (2.7)). Second, the claim of LL is lacking any scientific evidence, but this term they criticized is well-known in the respective literature (see, e.g., Liou[19]). Third, as it is shown in Appendix A, the term is correct. Fourth, if the atmosphere would serve as a perfect absorber, i.e., $a_a = \varepsilon_a = 1$, where $a_a$ is the integral absorptivity of the entire atmosphere in the infrared range, no long-wave (infrared) radiation emitted by the Earth's surface would, indeed, reach the space. Fifth, the reason for the temperature of $T_E = 268\,\mathrm{K}$ is an absorptivity of $A_a = 0.26$ (see Figure 15 of Ref. 2, reproduced here for convenience as Figure 3). This values of $A_a$ was already used in our Figure 14 (reproduced here for convenience as Figure 2). With respect to this figure we stated:[2]

> »If we additionally assume that also the Earth acts as a blackbody emitter, we will obtain a temperature value for the Earth's surface of $T_E = 288\ \mathrm{K}$ for an absorptivity of $A_a \cong 0.26$.
>
> This value is close to those estimated by Budyko [24], Schneider [17], Liou [18], and Trenberth et al. [29]….«

Note that the bibliography numbers in this quotation are related to the paper of KD. To understand these results, it is indispensable to briefly describe the Dines-type two-layer energy balance models. This will be done in the next subsection.

### 2.3 Dines-type two-layer energy balance models of the Earth-atmosphere system

As derived in the Appendix of KD, the basic equations for Dines-type two-layer energy balance model for the system Earth-atmosphere read:

Water layer of the aqua-planet:

$$R\,\frac{dT_s}{dt} = \left(1-\alpha_E - A_a\right)\frac{S}{4} - H - E + \varepsilon_E\ \varepsilon_a\ \sigma\ T_a^{\ 4} - \varepsilon_E\ \sigma\ T_s^{\ 4} \tag{2.12}$$

Atmosphere:

$$\left.\begin{aligned} R_a\,\frac{dT_a}{dt} &= \left(1-\alpha_E\right)\frac{S}{4} - \varepsilon_a\ \sigma\ T_a^{\ 4} - \left(1-\varepsilon_a\right)\varepsilon_E\ \sigma\ T_s^{\ 4} - \left(1-\alpha_E - A_a\right)\frac{S}{4} \\[2mm] &\quad - \varepsilon_E\ \varepsilon_a\ \sigma\ T_a^{\ 4} + \varepsilon_E\ \sigma\ T_s^{\ 4} + H + E \\[2mm] &= A_a\,\frac{S}{4} - \left(1+\varepsilon_E\right)\varepsilon_a\ \sigma\ T_a^{\ 4} + \varepsilon_a\ \varepsilon_E\ \sigma\ T_s^{\ 4} + H + E \end{aligned}\right\}\ . \tag{2.13}$$

Here, R and $R_a$ are so-called inertia coefficients, and $T_a$ and $\varepsilon_a$ are the temperature and the integral emissivity of the atmosphere, respectively. Equations (2.12) and (2.13) form a coupled system of ordinary differential equations. It can be solved numerically, where the solar constant, S, has to be considered as an external "force".

As shown in the instance of the water layer of an aqua-planet as considered by Schneider and Mass[3], these are integral energy balance equations for the entire layers under study. As documented by Eqs. (A6) and (A13) of KD, $T_s$ and $T_a$ are related to volume-averaged temperatures, where the volume average is defined by (see Eq. (A7) in Ref. 2)

$$\langle \Psi \rangle_V = \frac{1}{V_w} \int_{V_w} \Psi \, dV \quad . \tag{2.14}$$

To relate any surface temperature to such a volume-averaged temperature as done by Schneider and Mass[3] is, indeed, a strong simplification. It would only be correct in the special case that the temperature is homogeneously distributed in the water layer of the aqua-planet of a thickness of several decameters and the entire atmosphere, respectively. This is rather inappropriate in case of both the water layer and the entire atmosphere. A consequence of this simplification is that the amounts of $T_s$ and $T_a$ are quite different even though in the close vicinity of the Earth's surface the mean temperature of the atmosphere only slightly differs from the mean Earth's surface temperature under real world conditions. Because of these facts, the fluxes of sensible and latent heat, H and E, cannot be related to $T_s$ and $T_a$, as already discussed by KD (see Eqs. (72) and (73) of their paper). Hence, the quantities H and E have to be inserted into equation set (2.12) and (2.13), too. To relate these fluxes to the temperatures $T_s$ and $T_a$ as done by LL is in contradiction to the basic knowledge in micrometeorology.[21-24] We will further discuss this aspect in section 5.

For steady-state conditions as usually considered, Eqs. (2.12) and (2.13) provide

$$0 = \left(1 - \alpha_E - A_a\right)\frac{S}{4} - H - E + \varepsilon_E \, \varepsilon_a \, \sigma \, T_a^{\,4} - \varepsilon_E \, \sigma \, T_s^{\,4} \tag{2.15}$$

and

$$0 = A_a \, \frac{S}{4} - \left(1 + \varepsilon_E\right)\varepsilon_a \, \sigma \, T_a^{\,4} + \varepsilon_a \, \varepsilon_E \, \sigma \, T_s^{\,4} + H + E \quad . \tag{2.16}$$

Combining these two equations yields

$$0 = \left(1 - \alpha_E\right)\frac{S}{4} - \varepsilon_a \, \sigma \, T_a^{\,4} - \left(1 - \varepsilon_a\right)\varepsilon_E \, \sigma \, T_s^{\,4} \tag{2.17}$$

considered for the top of the atmosphere (TOA). Equation (2.15) is identical with Eq. (69) and Eq. (2.17) is identical with Eq. (64) of KD. Equation set (2.15) and (2.17) serves as the basis for Dines-type two-layer energy balance models of the Earth-atmosphere system[4]. In Eq. (2.17) the term $\left(1 - \varepsilon_a\right) \varepsilon_E \ \sigma \ T_s^{\ 4}$ occurs. As mentioned by KD, it describes the terrestrial radiation that is propagating through the atmosphere (it also includes the terrestrial radiation that is passing through the so-called atmospheric window), i.e., the term $1 - \varepsilon_a$ describes the integral transmissivity of the atmosphere in the infrared range.

Since formula (2.17) is a linear combination of Eqs. (2.15) and (2.16), they are only two coupled governing equations which are related to the boundaries of the integral balance equations. As the number of unknowns, $T_s$ and $T_a$, is equal to the number of governing equations, this equation set can be solved. If we assume for the purpose of simplification $\varepsilon_E = 1$ as done, for instance, by Kiehl and Trenberth[25] and Liou[19] we will obtain:

Earth's surface:

$$0 = \left(1 - \alpha_E - A_a\right) \frac{S}{4} - H - E + \varepsilon_a \ \sigma \ T_a^{\ 4} - \sigma \ T_s^{\ 4} \qquad (2.18)$$

Top of the Atmosphere:

$$0 = \left(1 - \alpha_E\right) \frac{S}{4} - \varepsilon_a \ \sigma \ T_a^{\ 4} - \left(1 - \varepsilon_a\right) \sigma \ T_s^{\ 4} \quad . \qquad (2.19)$$

The latter is identical with Liou's Eq. (8.3.3)[19]. If we neglect the fluxes of sensible and latent heat, H and E, in Eq. (18), we will lead to Liou's Eq. (8.3.4)[19]. LL completely disregarded these facts even though they cited Liou's textbook. It should be noticed that Link and Lüdecke[17] already claimed in their comment from December 2010 (in German):

>»Die Ergebnisse sind in den Abbildungen (14) und (15) graphisch dargestellt. In Glg. (64) ist ihnen dabei ein Fehler unterlaufen. Der erste Term muss heißen $A_a$·S/4 anstatt $(1 - \alpha_E)$·S/4.«

It means:

>The results are illustrated in Figures (14) and (15). In Eq. (64) they made a mistake. The first term must read $A_a$·S/4 instead of $(1 - \alpha_E)$·S/4.

This claim is wrong, too. Obviously, the statement of Link and Lüdecke[17] reflects that they were guessing only. Figure 2 of our reply to this comment[18] exhibits the page 461 of Liou's textbook[19], on which Liou's equation set is listed, i.e., LL knew that their claims are false, at least, since the beginning of January 2011, i.e., three months before they submitted the manuscript to the International Journal of Modern Physics C.

The solution of equation set (2.15) and (2.17) is given by:

$$T_a = \left\{ \frac{\left(A_a + \varepsilon_a \left(1 - \alpha_E - A_a\right)\right) \dfrac{S}{4} + \left(1 - \varepsilon_a\right)\left(H + E\right)}{\varepsilon_a \, \sigma \left(1 + \varepsilon_E \left(1 - \varepsilon_a\right)\right)} \right\}^{\frac{1}{4}} \tag{2.20}$$

and

$$T_s = \left\{ \frac{\left(\left(1 + \varepsilon_E\right)\left(1 - \alpha_E\right) - A_a\right) \dfrac{S}{4} - H - E}{\varepsilon_E \, \sigma \left(1 + \varepsilon_E \left(1 - \varepsilon_a\right)\right)} \right\}^{\frac{1}{4}} \quad . \tag{2.21}$$

If we ignore the fluxes of sensible and latent heat in this equation set for a moment, we will obtain

$$T_a = \left\{ \frac{\left(A_a + \varepsilon_a \left(1 - \alpha_E - A_a\right)\right) S}{4 \, \varepsilon_a \, \sigma \left(1 + \varepsilon_E \left(1 - \varepsilon_a\right)\right)} \right\}^{\frac{1}{4}} \tag{2.22}$$

and

$$T_s = \left\{ \frac{\left(\left(1 + \varepsilon_E\right)\left(1 - \alpha_E\right) - A_a\right) S}{4 \, \varepsilon_E \, \sigma \left(1 + \varepsilon_E \left(1 - \varepsilon_a\right)\right)} \right\}^{\frac{1}{4}} \quad . \tag{2.23}$$

For $\varepsilon_E = 1$ these equation set simplifies to

$$T_a = \left\{ \frac{\left(A_a + \varepsilon_a \left(1 - \alpha_E - A_a\right)\right) S}{4 \, \varepsilon_a \, \sigma \left(2 - \varepsilon_a\right)} \right\}^{\frac{1}{4}} \tag{2.24}$$

and

$$T_s = \left\{ \frac{\left(2 \left(1 - \alpha_E\right) - A_a\right) S}{4 \, \sigma \left(2 - \varepsilon_a\right)} \right\}^{\frac{1}{4}} \quad . \tag{2.25}$$

The formulae (2.24) and (2.25) are identical with Liou's[19] Eqs. (8.3.5) and (8.3.6) as already mentioned by KD. Again, LL disregarded these facts completely.

Figure 14 of KD (here repeated as Figure 2 for the purpose of convenience) is based on Eqs. (2.22) and (2.23). It is obvious that for $\varepsilon_a = 1$ the temperature $T_a$ becomes independent of the absorptivity, $A_a$, of the atmosphere in the solar range. This result also holds in case of the equation sets (2.20) and (2.21) as well as (2.24) and (2.25) because we always obtain

$$T_a = \left\{ \frac{\left( 1 - \alpha_E \right) S}{4 \, \sigma} \right\}^{\frac{1}{4}} \, . \tag{2.26}$$

Obviously, Eq. (2.26) must provide the same result like formula (46) of KD if $\varepsilon_E = 1$ and $\alpha_E = 0.3$ are chosen. Consequently, one obtains $T_a = 255 \, \text{K}$. This result is not odd, it is based on a simple algebra. From a physical point of view an emissivity of $\varepsilon_a = 1$ would mean that the atmosphere completely absorbs the radiation emitted by the Earth's surface. The temperature $T_a = 255 \, \text{K}$ is the temperature of the so-called planetary radiative equilibrium for the TOA. Assuming, for instance, $A_a = 0$, $\varepsilon_a = 1$, $\varepsilon_E = 1$, and, again, $\alpha_E = 0.3$ yields

$$T_s = \left\{ \frac{\left( 1 - \alpha_E \right) S}{2 \, \sigma} \right\}^{\frac{1}{4}} = 2^{\frac{1}{4}} \, T_a = 303 \, \text{K} \, . \tag{2.27}$$

Note that KD already stated:

»Assuming, for instance, that the atmosphere acts as blackbody emitter leads to an atmospheric temperature of about $T_a \cong 255 \, \text{K}$ which is independent of the absorptivity (see Eq. (66) for $\varepsilon_a = 1$). Considering, in addition, the Earth as a blackbody emitter provides a surface temperature of about $T_E \cong 303 \, \text{K}$ if $A_a$ is assumed to be zero. This value completely agrees with those of Smith [57], Hantel [58], and Kump et al. [59].«

Again, the bibliography numbers in this quotation are related to KD. Since the temperatures $T_a = 255 \, \text{K}$ and $T_s = 303 \, \text{K}$ are well known in the literature,[22, 26-29] the false claim of LL documents that these authors completely disregarded the respective literature in this matter, too.

If a value of $A_a = 0.26$ is considered in Figure 3, we will, indeed, obtain: $T_E = 268 \, \text{K}$. However, in contrast to the claim of LL, this temperature was not explicitly mentioned because the temperature $T_s$ shown in Figure 3 is generally lower than the corresponding one illustrated in Figure 2. The opposite is true in case of $T_a$. This is quite reasonable because the fluxes of sensible and latent heat considered in case of Figure 3 are directed upward and contribute to a "warming" of the atmosphere (correctly spoken: to an increase of the total internal energy) and a "cooling" of the (soil and) water layers adjacent to the Earth's surface. This behavior is documented by Eqs. (70) and (71) of our paper[2]. If we rearrange these equations and chose $\varepsilon_E = 1$ for the purpose of simplification we will obtain:

$$T_a = \underbrace{\left\{ \frac{\left( A_a + \varepsilon_a \left( 1 - \alpha_E - A_a \right) \right) S}{4 \, \varepsilon_a \, \sigma \left( 2 - \varepsilon_a \right)} \right\}^{\frac{1}{4}}}_{\text{Eq. (2.24)}} \left\{ 1 + \frac{4 \left( 1 - \varepsilon_a \right) \left( H + E \right)}{\left( A_a + \varepsilon_a \left( 1 - \alpha_E - A_a \right) \right) S} \right\}^{\frac{1}{4}} \qquad (2.28)$$

and

$$T_E = \underbrace{\left\{ \frac{\left( 2 \left( 1 - \alpha_E \right) - A_a \right) S}{4 \, \sigma \left( 2 - \varepsilon_a \right)} \right\}^{\frac{1}{4}}}_{\text{Eq. (2.25)}} \left\{ 1 - \frac{4 \left( H + E \right)}{\left( 2 \left( 1 - \alpha_E \right) - A_a \right) S} \right\}^{\frac{1}{4}} \qquad . \qquad (2.29)$$

Obviously, the second term on the right-hand side of Eq. (2.28) is greater than unity. Whereas the opposite is true in case of Eq. (2.29). Note that the globally averaged fluxes H and E are positive (see Figure 1).

### 2.4 The dependency of the atmospheric temperature $T_A$ on the atmospheric absorptivity aa in the solar range

As already mentioned before, LL falsely claimed:

»In their paper this value is independent of the integral absorptivity of the atmosphere with respect to the range of solar radiation ($A_a$ in Fig. 15 of Ref. 3), which seems rather odd.«:

They continued their false claim on page 453 by

»The mean atmospheric temperature $T_A$ as a function of aa can be evaluated from Eqs. (18) and (19). In a very good approximation the function is linear

$$T_A = 75.04 \cdot \text{aa} + 257 \left[ K \right]$$

Obviously the value for $T_A$ is strongly dependent on aa in disagreement with the results from Kramm and Dlugi, who calculated the mean temperature of the atmosphere as $T_A = 255 \, K$, independent of the integral absorption of the atmosphere with respect to the short wave radiation of the sun.«

As illustrated in Figures 14 and 15 of KD (reproduced here as Figures 2 and 3), in case of the integral emissivity of $\varepsilon_a < 1$, the temperature $T_a \left( = T_A \right)$, of course, depends on the atmospheric absorptivity in the solar range denoted by $A_a \left( = \text{aa} \right)$. As mentioned before, only in case of $\varepsilon_A = 1$ one obtains $T_a = 255$ K if $\alpha_E = 0.3$ is chosen (see Eq. (2.26)), i.e., the temperature of the so-called planetary radiative equilibrium at the TOA.

## 3. Falsified quotations

LL stated:

> »Kramm and Dlugi used the incompatibility of their model results with observed values as an argument that their model (the improved model of Schneider et al.) reveals "no evidence for the existence of the so-called atmospheric greenhouse effect, if realistic empirical data are used."«

This statement is based (a) on a false claim and (b) on a falsified quotation. Section 6 of our paper[2] in which two-layer energy balance model for the system Earth-atmosphere were analyzed begins as follows (the bibliography numbers in this quotation are related to KD):

> »Recently, Smith [57] discussed the IR absorption by the atmosphere to illustrate the so-called greenhouse effect, where he used a two-layer model of radiative equilibrium. Similar models were already discussed, for instance, by Hantel [58] and Kump et al. [59]. In contrast to these models in which the absorption of solar radiation by the atmosphere is not included we consider the more advanced one of Dines [60] (see Figure 13) and Liou [18].«

These two sentences point out that our intension was to analyze two-layer energy balance models published in the literature. To claim that the Dines-type two-layer model assessed by us is our improved model of Schneider and Mass[3] is, therefore, unabashed because the model structure of Dines[4] is 58 years older than that of Schneider and Mass.[3] Note that the section "Introduction" by Kiehl and Trenberth[25] begins as follows:

> »There is a long history of attempts to construct a global annual mean surface–atmosphere energy budget for the earth. The first such budget was provided by Dines (1917).«

Since LL cited the paper of Kiehl and Trenberth, they should know that Dines-type energy balance models have a long history.

In contrast to the falsified quotation of LL, we already stated in the "Abstract" of our paper:[2]

> »Moreover, both the model of Schneider and Mass and the Dines-type two-layer energy balance model for the Earth-atmosphere system, containing the planetary radiation balance for the Earth in the absence of an atmosphere as an asymptotic solution, do not provide evidence for the existence of the so-called atmospheric greenhouse effect if realistic empirical data are used.«

In Section 7 of our paper we concluded:[2]

> »Based on our findings we may conclude that it is time to acknowledge that the principles on which global energy balance climate models like that of Schneider and Mass [1] or that of Dines [2] are based have serious physical shortcomings and should not further be used to study feedback mechanisms, transient climate response, and the so-called atmospheric greenhouse effect.«

Again, the bibliography numbers in this quotation are related to KD.

4. **The new basic 1-dimensional 1-layer model of LL**

First at all, the new basic 1-dimensional 1-layer model of LL is neither a one-dimensional nor a one-layer model because (a) there is no spatial coordinate (e.g., the vertical coordinate) and (b) it contains two layers: one layer beneath the Earth's surface, i.e., the layer of the aqua-planet (see Eq. (2.12)), and one layer above the Earth's surface, i.e., the atmosphere (see Eq. (2.13)). The structure of such two-layer energy balance models is sketched in Figure 13 of KD here reproduced as Figure 4. This figure is taken from Dines' paper.[4] Note that in the following sections we use the symbols of LL.

*4.1 Confused handling of the solar radiation absorbed by the Earth by LL*
In their paper LL defined the solar radiation absorbed by the Earth by (see also Appendix B and Figure 1)

$$S_E = \left(1 - a - aa\right)\frac{S_0}{4} \quad . \tag{4.1}$$

This means that $S_{ER}$ must not occur in Eq. (4) of LL. In contrast to formula Eq. (9) of LL, this mistake leads to

$$S_E = \left(1 - a - aa\right)\frac{S_0}{4} + S_{ER} \quad . \tag{4.2}$$

Inserting, for instance, the data of Trenberth et al.[12] would yield: $S_E = 184 \text{ W m}^{-2}$. This value notably disagrees with the value of $S_E = 161 \text{ W m}^{-2}$ shown in Figure 1. If we consider the Earth in absence of its atmosphere, Eq. (31) would lead to

$$S_E = \left(1 - \frac{4\,S_{ER}}{S_0}\right)\frac{S_0}{4} + S_{ER} = \frac{S_0}{4} = 341 \text{ W m}^{-2} \quad . \tag{4.3}$$

In such a case the temperature of the planetary radiation balance of the Earth in the absence of the atmosphere would be 278 K, instead of the commonly obtained 255 K (if $\varepsilon_E = 1$ and $\alpha_E = 0.3$ are chosen). Therefore, to reduce $S_E$ by the reflected one, $S_{ER}$, is rather inadequate because in such a case $S_E$ cannot be the absorbed solar radiation. Note that these two mistakes were cured in their Eq. (16) because they inserted their improper Eq. (9) into their improper Eq. (4).

### 4.2 Reflection of long-wave radiation emitted by the Earth's surface by the atmosphere

LL re-established the reflection of long-wave radiation - emitted by the Earth's surface - by the atmosphere. As illustrated in Figure 4, such a term was already recommended by Dines.[4] As mentioned by KD, Möller[30] criticized the Dines' representation because the reflection of long-wave radiation in the atmosphere plays no role in this case, with exception of scattering of infrared radiation on small ice particles of cirrus clouds. This scattering of infrared radiation by cirrus clouds plays a notable role in the so-called atmospheric window at both sides of the $9.6\,\mu m$ -band of ozone. These scattering processes cannot be handled in such two-layer energy balance models for the Earth-atmosphere system, neither forward nor backward scattering. We did not only cite Möller[30], but we also stated:[2]

> »Moreover, the reflection of IR radiation at the Earth's surface is included here, but scattering of IR radiation in the atmosphere is ignored, in accord with Möller [61]. The latter substantially agrees with the fact that in the radiative transfer equation the Planck function is considered as the only source function when a non-scattering medium is in local thermodynamic equilibrium so that a beam of monochromatic intensity passing through the medium will undergo absorption and emission processes simultaneously, as described by Schwarzschild's equation [18, 63-65].«

Again, the bibliography numbers in this quotation are related to KD. The model completeness claimed by LL is, therefore, based on their attempt to re-establish a term considered as rather unimportant in the literature.

Since LL considered the paper of Trenberth et al.[12] as their reference, they did not find any energy flux related to the long-wave radiation emitted by the Earth's surface and reflected by the atmosphere. LL assumed, therefore, that $r_{EA} = 1 - \varepsilon_A = 0$ leading to $\varepsilon_A = 1$. This is clearly wrong. Because of the conservation of radiation propagating through the atmosphere we have (see Appendix A)

$$r_{EA} + a_A + d = 1 \quad . \tag{4.4}$$

Here, $r_{EA}$ is the integral reflectivity, $a_A$ is the integral absorptivity, and $d$ is the integral transmissivity, where $a_A = \varepsilon_A$ is customarily used. This means that in case of $\varepsilon_A = 1$ both $r_{EA}$ and $d$ are equal to zero because $r_{EA} \geq 0$ and $d \geq 0$. Thus, in case of $\varepsilon_A = 1$ the long-wave radiation emitted by the Earth's surface would completely be absorbed, i.e., a direct propagation of long-wave radiation from the Earth's surface into the space would not be possible. This fact is expressed by the term

$$\left(1 - \varepsilon_a\right) \varepsilon_E \, \sigma \, T_s^4 \quad , \tag{4.5}$$

strongly criticized by LL because of unknown reasons (see subsection 2.2 of this comment), where the term $1 - \varepsilon_a$ expresses the integral transmissivity of the entire atmosphere.

### 4.3 The boundary conditions at the top of the atmosphere

Beside the reflection of infrared radiation in the atmosphere, LL also considered an infrared flux term $L_{ToA}$ for the TOA. Using the power law of Stefan and Boltzmann they related a temperature $T_{ToA}$ to this flux term like in case of the Earth's surface (see their Eq. (21)). This flux term also occurs in any other two-layer Dines-type model and even in Liou's model. This will be shown here. Equations (14) and (16) of LL read:

$$0 = (1-a)\frac{S_0}{4} - L_{ToA} - d\,L_E \tag{4.6}$$

and

$$0 = (1-a-aa)\frac{S_0}{4} - L_E\left(1 - r_{EA} - d\,r_{EA}\right) + L_A\left(1 - r_A\right) - Q \quad . \tag{4.7}$$

Rearranging these equations yields

$$L_E = \frac{1}{d}\left((1-a)\frac{S_0}{4} - L_{ToA}\right) \tag{4.8}$$

and

$$L_A = \frac{1}{1-r_A}\left(L_E\left(1 - r_{EA}\left(1+d\right)\right) - \left(1-a-aa\right)\frac{S_0}{4} + Q\right) \quad . \tag{4.9}$$

These equations are identical with Eqs. (17) and (18) of LL. Even they claimed that the reflection of long-wave radiation by the atmosphere as important, which is in contradiction of the literature, LL did not further consider this reflection, i.e., $r_{EA} = 0$, and since $a_E = \varepsilon_E = 1 - r_A$. Thus, Eq. (4.7) may be written as

$$0 = (1-a-aa)\frac{S_0}{4} - L_E + \varepsilon_E\,L_A - Q \quad . \tag{4.10}$$

LL expressed the fluxes $L_A$ and $L_E$ by their Eqs. (19) and (20), respectively. Thus, Eqs. (4.6) and (4.10) may be expressed by

$$0 = (1-a)\frac{S_0}{4} - L_{ToA} - \left(1 - \varepsilon_A\right)\varepsilon_E\,\sigma\,T_E^{\ 4} \tag{4.11}$$

and

$$0 = (1-a-aa)\frac{S_0}{4} - \varepsilon_E\,\sigma\,T_E^{\ 4} + \varepsilon_E\,\varepsilon_A\,\sigma\,T_A^{\ 4} - Q \quad , \tag{4.12}$$

where – in accord with Eq. (4.4) – the integral transmissivity, d, in Eq. (4.6) has been replaced by $1 - \varepsilon_A$ because $r_{EA} = 0$. Note that Eq. (4.12) completely agrees with Eq. (69) of KD listed here as Eq. (2.15).

If we assume that the emission of infrared radiation in the atmosphere is isotropic, the flux term $L_{ToA}$ can be expressed by[19]

$$L_{ToA} = \varepsilon_A \, \sigma \, T_A{}^4 \quad . \tag{4.13}$$

Thus, Eq. (4.11) can be written as

$$0 = (1 - a)\frac{S_0}{4} - \varepsilon_A \, \sigma \, T_A{}^4 - (1 - \varepsilon_A)\, \varepsilon_E \, \sigma \, T_E{}^4 \quad . \tag{4.14}$$

This equation completely agrees with Eq. (64) of KD listed here as Eq. (2.17). The solution of the equation set (2.15) and (2.17) is given by Eqs. (2.20) and (2.21) in subsection 2.3.

On the other hand, using Eq. (21) of LL leads to

$$0 = (1 - a)\frac{S_0}{4} - \varepsilon_{ToA} \, \sigma \, T_{ToA} - (1 - \varepsilon_A)\, \varepsilon_E \, \sigma \, T_E{}^4 \quad . \tag{4.15}$$

The solution of the equation set (4.12) and (4.15) reads:

$$T_A = \left\{ \frac{\left(aa + \varepsilon_A \left(1 - a - aa\right)\right)\frac{S_0}{4} - \varepsilon_{ToA} \, \sigma \, T_{ToA}{}^4 + (1 - \varepsilon_A)\, Q}{(1 - \varepsilon_A)\, \varepsilon_E \, \varepsilon_A \, \sigma} \right\}^{\frac{1}{4}} \tag{4.16}$$

and

$$T_E = \left\{ \frac{(1 - a)\frac{S_0}{4} - \varepsilon_{ToA} \, \sigma \, T_{ToA}{}^4}{(1 - \varepsilon_A)\, \varepsilon_E \, \sigma} \right\}^{\frac{1}{4}} \quad . \tag{4.17}$$

This means that this equation set contains three unknown, namely $T_A$, $T_E$, and $T_{ToA}$, but only two governing equations are available. This equation set cannot directly be solved. Further assumptions regarding one of these three unknowns are indispensable to achieve a tractable equation set. Setting $\varepsilon_{ToA} = 1$ as done by LL leads to another model inconsistency.

To express the flux term $L_{ToA}$ by the temperature $T_{ToA}$ at the TOA is physically inappropriate. The flux term $L_{ToA}$ is caused by a layer of the atmosphere, but not by a surface. Unfortunately, LL completely ignored in their model description that the emitted infrared radiation is an example of isotropic radiation. On the one hand, there is no downward directed flux that would affect the atmosphere beneath the layer which emits $L_{ToA}$. On the other hand, there is no upward directed flux $L_A$ that would affect the atmospheric layer above the layer which emits $L_A$. If the prerequisite of isotropy is ignored the dependence of Stefan's constant, $\sigma = 2\,\pi^5\,k^4 / (15\,c^2\,h^3) \cong 5{,}67 \cdot 10^{-8}\ \mathrm{W\ m^{-2}\ K^{-4}}$, on $\pi^5$ would not generally exist because the derivation of the power law of Stefan and Boltzmann is not only based on the integration of Planck's radiation function over all frequencies (alternatively all wavelengths or all wave numbers) which leads to $I = b\,T^4$ with $b = 2\,\pi^4\,k^4 / (15\,c^2\,h^3)$, but also the integration of the total intensity, $I$, over the adjacent halve space which provides $L = \pi\,I$ if the assumption of isotropic radiation is considered.[19, 27, 31] Note that $k = 1{,}381 \cdot 10^{-23}\ \mathrm{J\ K^{-1}}$ is Boltmann's constant, $h = 6{,}626 \cdot 10^{-34}\ \mathrm{J\ s}$ is Planck's constant, and $c = 2{,}998 \cdot 10^8\ \mathrm{m\ s^{-1}}$ is the velocity of light in vacuum. Furthermore, at altitudes of more than 60 km or so the assumption of local thermodynamic equilibrium is not valid. Thus, the source function in the radiative transfer equation should not be substituted by Planck's radiation function. This means that using the power law of Stefan and Boltzmann for the TOA is not straight-forward.

## 5. Assessing the model results of LL

LL did not derive equation set (4.16) and (4.17). Thus, the results for the temperatures $T_A$ and $T_E$ given by LL on page 453 are not based on their model predictions using this (or a similar) equation set, but on the estimated fluxes $L_A = 333\ \mathrm{W\ m^{-2}}$ and $L_E = 396\ \mathrm{W\ m^{-2}}$ of Trenberth et al.[12] that are illustrated here in Figure 1. This means that LL only expressed $L_A$ and $L_E$ by the power law of Stefan and Boltzmann, where they assumed $\varepsilon_A = \varepsilon_E = 1$, i.e., they inversely applied this power law. Thus, they obtained: $T_A = 277\ \mathrm{K}$ and $T_E = 289\ \mathrm{K}$. The former has no relevance as a representative (volume-averaged) temperature for the lower atmosphere (as mentioned by KD, the mean temperature of the troposphere amounts to 255 K), and the latter is close to the globally averaged temperature which was determined by Trenberth et al.[12] on the basis of the surface skin temperature provided by the NCEP-NCAR reanalysis at T62 resolution. This means that the excellent agreement of the model results claimed by LL is based on the fact that they reproduced the input of Trenberth et al.[12] on which $L_E = 396\ \mathrm{W\ m^{-2}}$ is based.

### 5.1 Figure 2 of LL

Figure 2 of LL is based on their equations (17) and (18) (here listed as Eqs. (4.8) and (4.9)), where they again assumed $r_{EA} = 1 - \varepsilon_A = 0$, i.e., $\varepsilon_A = 1$, and $r_A = 1 - \varepsilon_E = 0$, i.e., $\varepsilon_E = 1$. LL pointed out that these assumptions are in agreement with Kiehl and Trenberth. This is not entirely correct because the assumption $\varepsilon_A = 1$ includes that the integral transmissivity of the respective atmospheric layer is equal to zero, but the energy flux schemes of both Kiehl and Trenberth and Trenberth et al. suggest that a portion of infrared radiation of about $40\ \mathrm{W\ m^{-2}}$ is traversing the entire atmosphere from the Earth's surface to the space. This is impossible if an atmospheric layer has an integral emissivity which is equal to unity.

As illustrated in the Figure 2 of LL and our Figure 5, $L_E$ and $L_A$ are functions of $1/d$ (see Eqs. (17) and (18) of LL), i.e., they are hyperbolas. Thus, $L_E$ and $L_A$ tend to infinity if d tends to zero. The same is true in case of the corresponding temperatures $T_E$ and $T_A$. This behavior is in complete disagreement with the results that can be found in the literature.[22, 26-29] As shown before (see Eqs. (2.26) and (2.27)) $T_E$ and $T_A$ are finite. These awkward results of LL are a consequence of the distinct inconsistency inherent in their new model.

### 5.2 Figure 3 of LL

LL stated that their Figure 3 is based on their Eqs. (17) to (20). If this is true, then their results illustrated in this figure are completely wrong. Obviously, the fluxes of sensible ($Q_S$) and latent heat ($Q_L$) do not occur in Eq. (17) of LL (here listed as Eq. (4.8)). This means that the flux $L_E$ is not affected by $Q = Q_S + Q_L$. Consequently, the corresponding temperature $T_E$ related to $L_E$ by Eq. (20) of LL must be constant for any value of $Q$. This behavior is illustrated in our Figure 7. According Eq. (18) of LL, the fluxes of sensible and latent heat only affect $L_A$ and, hence, $T_A$. Also this behavior is shown in our Figure 7. In the capture of their Figure 3 LL pointed out that $T_A = 227\,K$ and $T_E = 271\,K$ leads to $Q = 7\,W\,m^{-2}$. Thus, the net radiation in the infrared range amounts to $\Delta F_{IR} = 155\,W\,m^{-2}$. This value completely disagrees with $\Delta F_{IR} = 63\,W\,m^{-2}$ of Trenberth et al.[12] We wonder whether this is the excellent agreement that was claimed by LL.

Beside these completely awkward results, LL tried to link the fluxes of sensible and latent heat to the globally averaged temperatures like $T_A$ and $T_E$. There is no physical law that describes the relationship of the fluxes of sensible and latent heat to such volume-averaged quantities. If we consider the vertical components of the fluxes of sensible and latent heat, we will obtain

$$H = -\overline{\rho}\,c_p\left(\alpha_T\frac{\partial\overline{T}}{\partial z} - \overline{w'\Theta'}\right) \tag{5.1}$$

and

$$E = -\overline{\rho}\,L_V\left(D_q\frac{\partial\overline{q}}{\partial z} - \overline{w'q'}\right)\quad, \tag{5.2}$$

where the vertical components of the respective gradients characterize the molecular effects (in accord with the laws of Fourier and Fick, respectively) and the covariance terms represent the turbulent effects. Here, $\alpha_T$ is the thermal diffusivity, $D_q$ the diffusivity of water vapor in air, $c_p$ is the specific heat at constant pressure, $L_v$ is, again, the specific heat of phase transition, w is the vertical component of the wind vector, $\Theta$ is the potential temperature, and q is the specific humidity. In the sense of Reynolds, the overbar characterizes the mean value and the prime the departure from that.

From a theoretical point of view the fluxes of sensible and latent heat are called non-convective fluxes, sometimes also called sub-scale fluxes. These fluxes are usually parameterized as expressed by Eqs. (72) and (73) of our paper.[2] This parameterization is in substantial agreement with the micrometeorological literature,[21-24] completely ignored by LL.

### 5.3 Climate sensitivity

LL stated:

> »As a result, the change of the surface temperature of the Earth induced by 3.7W/m$^2$ radiative forcing (doubling of $CO_2$) is $\Delta T \approx 1.1\,\mathrm{K}$, which is in good agreement with the IPCC value (without feedback).«

Beside the fact that this result is not based on the model of LL, it completely disagrees with the definition of the radiative forcing, RF, as considered in the 4[th] report Working Group I to the Intergovernmental Panel on Climate Change (IPCC), Climate Change 2007 – The Physical Science Basis. An excerpt of subsection 2.2 of this report reads:[32]

> »The definition of RF from the TAR and earlier IPCC assessment reports is retained. Ramaswamy et al. (2001) define it as 'the change in net (down minus up) irradiance (solar plus longwave; in $\mathrm{W\,m^{-2}}$) at the tropopause after allowing for stratospheric temperatures to readjust to radiative equilibrium, but with surface and tropospheric temperatures and state held fixed at the unperturbed values'. ….. Radiative forcing can be related through a linear relationship to the global mean equilibrium temperature change at the surface $(\Delta T_s)$: $\Delta T_s = \lambda * RF$, where $\lambda *$ is the climate sensitivity parameter (e.g., Ramaswamy et al., 2001).«

As discussed by KD in their section 2, the quantity $\Delta T_s = \lambda * RF$ is related to either Eqs. (2.3) and (2.4), where $Q$ is expressed by

$$Q = \underbrace{\left(1 - \alpha_E\right)\frac{S}{4} - a + b\,T_r}_{Eq.\,(2.4)} + RF \quad , \tag{5.3}$$

or to Eqs. (2.3) and (2.7), where $Q$ is given by

$$Q = \underbrace{\left(1 - \alpha_E - A_a\right)\frac{S}{4} - H - E - a + b\,T_r}_{Eq.\,(2.7)} + RF \quad . \tag{5.4}$$

Solving Eq. (2.3) for the undisturbed system and the system disturbed by the anthropogenic radiative forcing alternatively, one obtains:

$$\Delta T_s = \lambda\,T_s^{(d)} - \lambda\,T_s = Q^{(d)} - Q = RF \tag{5.5}$$

or

$$\Delta T_s = T_s^{(d)} - T_s = \frac{Q^{(d)}}{\lambda} - \frac{Q}{\lambda} = \frac{RF}{\lambda} = \lambda * RF \qquad (5.6)$$

because the climate sensitivity parameter is the reciprocal of the feedback parameter. Note that $(d)$ marks the solution of the disturbed one. Thus, LL compared apples with plums.

## 6. Final remarks and conclusions

In our comments on the paper of Link and Lüdecke we documented that LL disobeyed scientific standards in various manner. Obviously, LL completely disregarded the literature cited not only in our paper, but also in their paper because many of their claims disagree with it. Furthermore, we showed that LL used many false claims and improper quotations in their attempt to blame us for the physical weakness of the theoretical concepts used in climatology like the global energy model of Schneider and Mass[3] and the Dines-type two-layer energy balance model for the Earth-atmosphere system.[4] This attempt of LL is unabashed because we concluded that the physical weakness is the reason why these theoretical concepts do not provide evidence for the existence of the so-called atmospheric greenhouse effect if realistic empirical data are used.

In our comments we also documented that the paper of LL is based on various fundamental physical and mathematical mistakes. Obviously, the new model of LL which is neither a one dimensional nor a one-layer model contains distinct inconsistencies that cause results which completely disagree with those that can be found in the literature. LL, for instance, ignored that infrared radiation is an example of isotropic radiation, a fundamental prerequisite in deriving the power law of Stefan and Boltzmann. Therefore, it is not surprising for us that LL obtained values for $L_E$, $L_A$, $T_E$, and $T_A$ that tend to infinity if the transmissivity of the atmosphere, $d$, tends to zero. This is in complete contradiction to the finite results for $L_E$ and $L_A$, $T_E$, and $T_A$ listed in the literature.[22, 26-29] Furthermore, we also documented that the results illustrated in Figure 3 of LL are completely wrong.

LL claimed that their result for the increase in the Earth's surface temperature of $\Delta T \approx 1.1\,K$ due to the doubling of the atmospheric $CO_2$ concentration is in good agreement with the IPCC value if no feedback is considered. However, beside the fact that this results is not based on their model predictions, it also disagrees with the definition of the anthropogenic radiative forcing as can be found not only in the respective literature, but also in various IPCC reports.[32] Thus, LL compared apples with plums.

The results for the temperatures $T_A$ and $T_E$ listed by LL on page 453 are not based on their model predictions, but on the estimated fluxes $L_A = 333\,W\,m^{-2}$ and $L_E = 396\,W\,m^{-2}$ of Trenberth et al.[12] illustrated here in Figure 1. This means that LL only expressed $L_A$ and $L_E$ by the power law of Stefan and Boltzmann, where they assumed $\varepsilon_A = \varepsilon_E = 1$. Thus, they obtained: $T_A = 277\,K$ and $T_E = 289\,K$. As mentioned before, the former has no relevance as a representative (volume-averaged) temperature for the lower atmosphere, and the latter is close to the globally averaged temperature which was determined by Trenberth et al.[12] on the basis of the surface skin temperature provided by the NCEP-NCAR reanalysis at T62 resolution. This means that the excellent agreement of the model results claimed by LL is based on the fact that they reproduced the input of Trenberth et al.[12] on which $L_E = 396\,W\,m^{-2}$ is based.

It should be noticed that Bruce B. Hicks, former Director of the Air Resources Laboratory of the National Oceanic and Atmospheric Administration (NOAA) of the United States of America stated in a book review two decades ago:[33]

»Some papers report on models that are developed, adjusted to fit a set of observations, and then 'validated' by comparison against the same data set. As a scientific community, we should urgently work on setting an appropriate punishment for those convicted of this crime.«

## Appendix A: Transmission of infrared radiation across the atmosphere

The monochromatic (or spectral) transmissivity (also called the transmittance) for the entire path through the atmosphere is defined by

$$\tau_\lambda = \frac{I_{\lambda,\mathrm{TOA}}}{I_{\lambda,\mathrm{E}}} \tag{A1}$$

Here, $I_{\lambda,\mathrm{TOA}}$ and $I_{\lambda,\mathrm{E}}$ are the monochromatic intensities of radiation at the top of the atmosphere (subscript TOA) and at the Earth's surface (subscript E), respectively. The monochromatic transmissivity fulfills the condition: $0 \le \tau_\lambda \le 1$. This means:

$$\tau_\lambda \begin{cases} = 1 & \text{there is no attenuation by absorption and / or reflection} \\ < 1 & \text{there is attenuation by absorption and / or reflection} \\ = 0 & \text{there is no transmission} \end{cases}$$

Be $I_{\lambda,\mathrm{a}}$ and $I_{\lambda,\mathrm{r}}$ monochromatic intensities of radiation that are absorbed (subscript a) and reflected (or back-scattered; subscript r) on the path through the entire atmosphere, we may define the monochromatic absorptivity and the monochromatic reflectivity by

$$a_\lambda = \frac{I_{\lambda,\mathrm{a}}}{I_{\lambda,\mathrm{E}}} \tag{A2}$$

and

$$r_\lambda = \frac{I_{\lambda,\mathrm{r}}}{I_{\lambda,\mathrm{E}}} \quad . \tag{A3}$$

Since

$$I_{\lambda,\mathrm{E}} = I_{\lambda,\mathrm{TOA}} + I_{\lambda,\mathrm{a}} + I_{\lambda,\mathrm{r}} \quad , \tag{A4}$$

we may write:[19, 34]

$$1 = \tau_\lambda + a_\lambda + r_\lambda \quad . \tag{A5}$$

Scatter processes in the infrared range, however, play only a minor role. (It has to be incorporated in case of ice clouds.) We already mentioned this fact in our paper. Thus, we may write

$$1 = \tau_\lambda + a_\lambda \tag{A6}$$

or

$$\tau_\lambda = 1 - a_\lambda \quad , \tag{A7}$$

With respect to Kirchhoff's law, the monochromatic absorptivity is related to the relative emissivity (defined as the ratio of the emitted intensity to the Planck function[19], $\varepsilon_\lambda$, by

$$\varepsilon_\lambda = a_\lambda \quad . \tag{A8}$$

Note that the use of Kirchhoff's law and the Planck function requires that the condition of local thermodynamic equilibrium (LTE) is fulfilled (in the atmosphere given up to a height above sea level of nearly 60 km). Thus, Eq. (A7) can also be expressed by

$$\tau_\lambda = 1 - \varepsilon_\lambda \quad . \tag{A9}$$

Using this formula in Eq. (A1) leads to

$$I_{\lambda, TOA} = \left(1 - \varepsilon_\lambda\right) I_{\lambda, E} \quad . \tag{A10}$$

The integration over all wavelengths provides

$$I_{TOA} = \int_0^\infty I_{\lambda, TOA} \, d\lambda = \int_0^\infty \left(1 - \varepsilon_\lambda\right) I_{\lambda, E} \, d\lambda = I_E - \int_0^\infty \varepsilon_\lambda \, I_{\lambda, E} \, d\lambda \quad , \tag{A11}$$

where

$$\int_0^\infty \varepsilon_\lambda \, I_{\lambda, E} \, d\lambda = \sum_i \int_{\Delta_i \lambda} \varepsilon_\lambda \, I_{\lambda, E} \, d\lambda \tag{A12}$$

expresses that the absorptivity/emissivity is related to certain bands (or even lines). If we define a mean emissivity (or, alternatively, the mean absorptivity) by

$$\varepsilon_a = \frac{\int_0^\infty \varepsilon_\lambda \, I_{\lambda,E} \, d\lambda}{\int_0^\infty I_{\lambda,E} \, d\lambda} = \frac{1}{I_E} \int_0^\infty \varepsilon_\lambda \, I_{\lambda,E} \, d\lambda \qquad (A13)$$

we will obtain:

$$I_{TOA} = \left(1 - \varepsilon_a\right) I_E \quad . \qquad (A14)$$

Since the emitted irradiance at the Earth's surface, $F_{IR\uparrow}$, differs from the emitted intensity, $I_E$, only by the factor of $\pi$ (related to the integration of the intensity over the adjacent half space[19, 31]), we may express the irradiance at the TOA by

$$F_{IR,TOA} = \left(1 - \varepsilon_a\right) F_{IR\uparrow} \quad . \qquad (A15)$$

Note that in case of the down-welling infrared radiation, $F_{IR\downarrow}$, the transmission at the Earth's surface would be negligibly small, but a notable portion would be reflected. Thus, instead of Eq. (A6) we have to consider

$$1 = a_\lambda + r_\lambda \quad . \qquad (A16)$$

Analogous to the procedure described by Eqs. (A6) to (A15), the part of the down-welling infrared radiation absorbed at the Earth's surface is given by[35]

$$a_E \, F_{IR\downarrow} = \left(1 - r_E\right) F_{IR\downarrow} = \varepsilon_E \, F_{IR\downarrow} \quad . \qquad (A17)$$

Thus, the reflectivity at the Earth's surface can be expressed by $r_E = 1 - \varepsilon_E$.

If we adopt the common procedure to express the emission by the Earth's surface in Dines-type two-layer energy balance models of the Earth-atmosphere system by one surface temperature, $T_E$, and recognize that the Earth's surface is a gray emitter, we will obtain:

$$F_{IR,TOA} = \left(1 - \varepsilon_a\right) \varepsilon_E \, \sigma \, T_E^{\,4} \quad . \qquad (A18)$$

This is the expression used in Eqs. (2.13) and (2.17). Consequently, the claim of LL is clearly wrong.

## Appendix B: Absorption of solar radiation

In case of solar radiation Eq. (A5) has to be applied. Thus, in two-layer energy balance models, the solar irradiance absorbed by the layer immediately beneath the Earth's surface is considered by[2, 19]

$$F_{S,E} = \left(1 - \alpha_E - A_a\right)\frac{S}{4} \quad . \tag{B1}$$

This expression was used in Eqs. (2.12) and (2.13). If we consider, for instance, the values of Trenberth et al.[12] here illustrated in Figure 1, we will obtain: $F_{S,E} = 161 \text{ W m}^{-2}$, where $\alpha_E = \left(79 + 23\right)/341 \cong 0.30$ and $A_a = 78/341 \cong 0.23$.

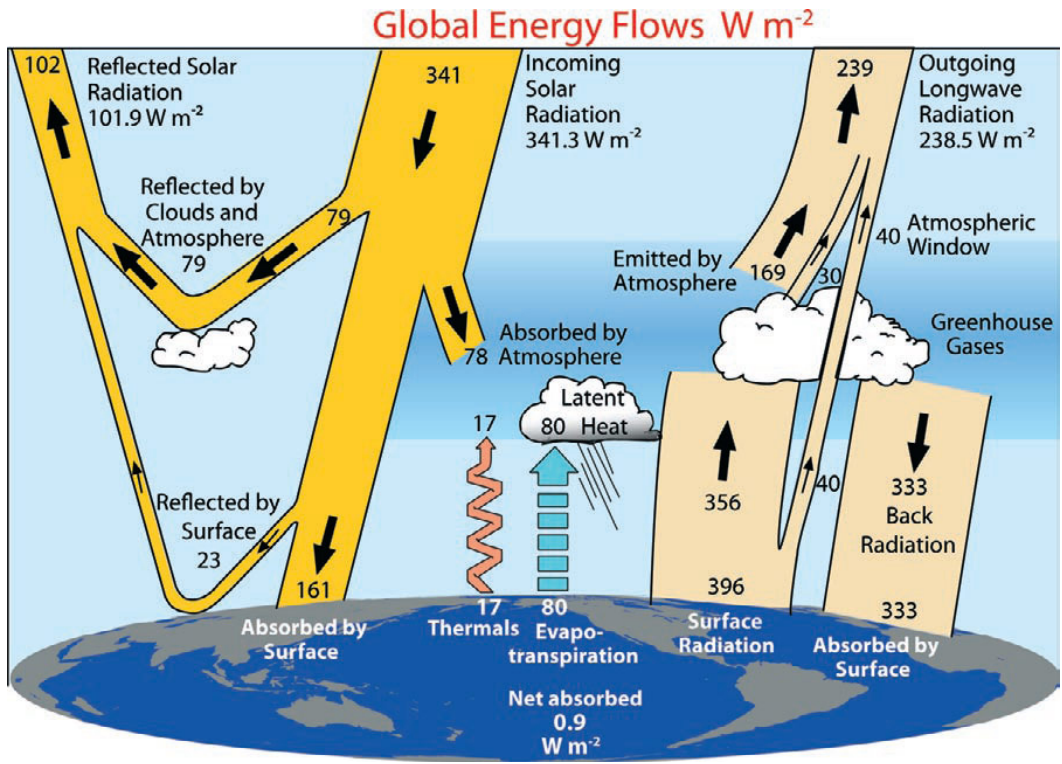

**Figure 1**: The global annual mean Earth's energy budget for the Mar 2000 to May 2004 period (W m$^{-2}$). The broad arrows indicate the schematic flow of energy in proportion to their importance (adopted from Trenberth et al.[12]).

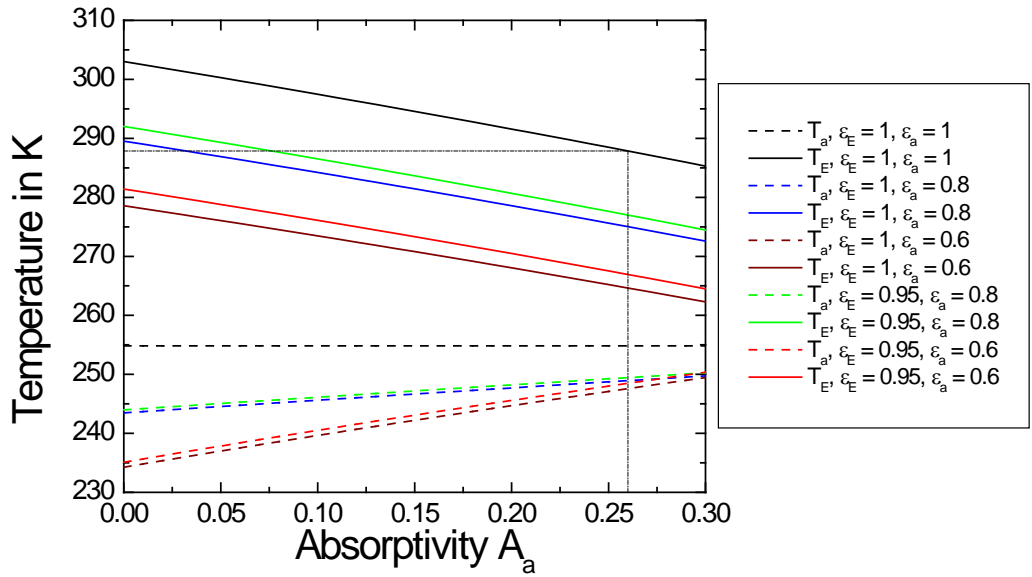

**Figure 2:** Uniform temperatures for the Earth's surface and the atmosphere provided by the two-layer model of a global energy-flux budget versus absorptivity $A_a$ (adopted from Refs. 2 and 20).

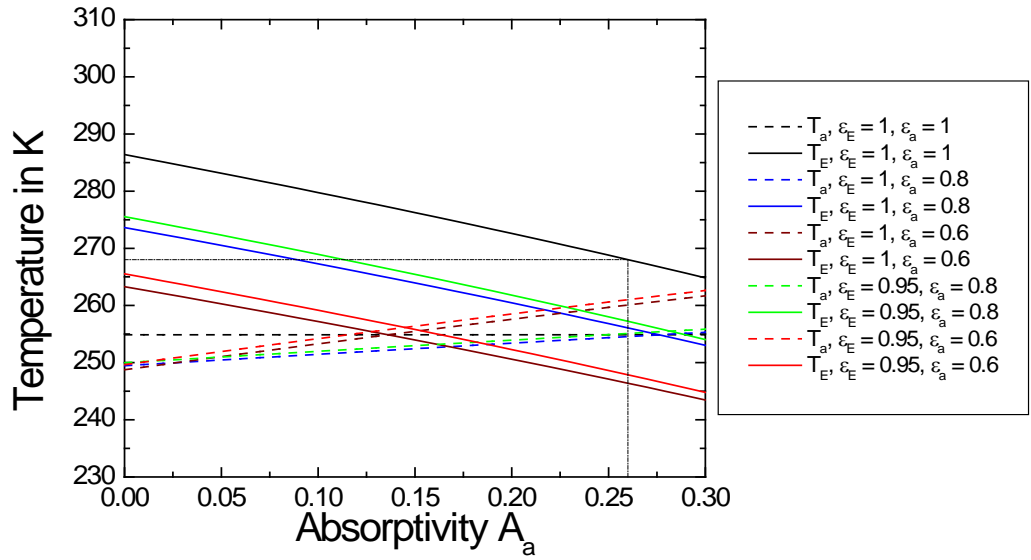

**Figure 3:** As in Figure 2, but the fluxes of sensible and latent heat are included (adopted from refs. 2 and 20).

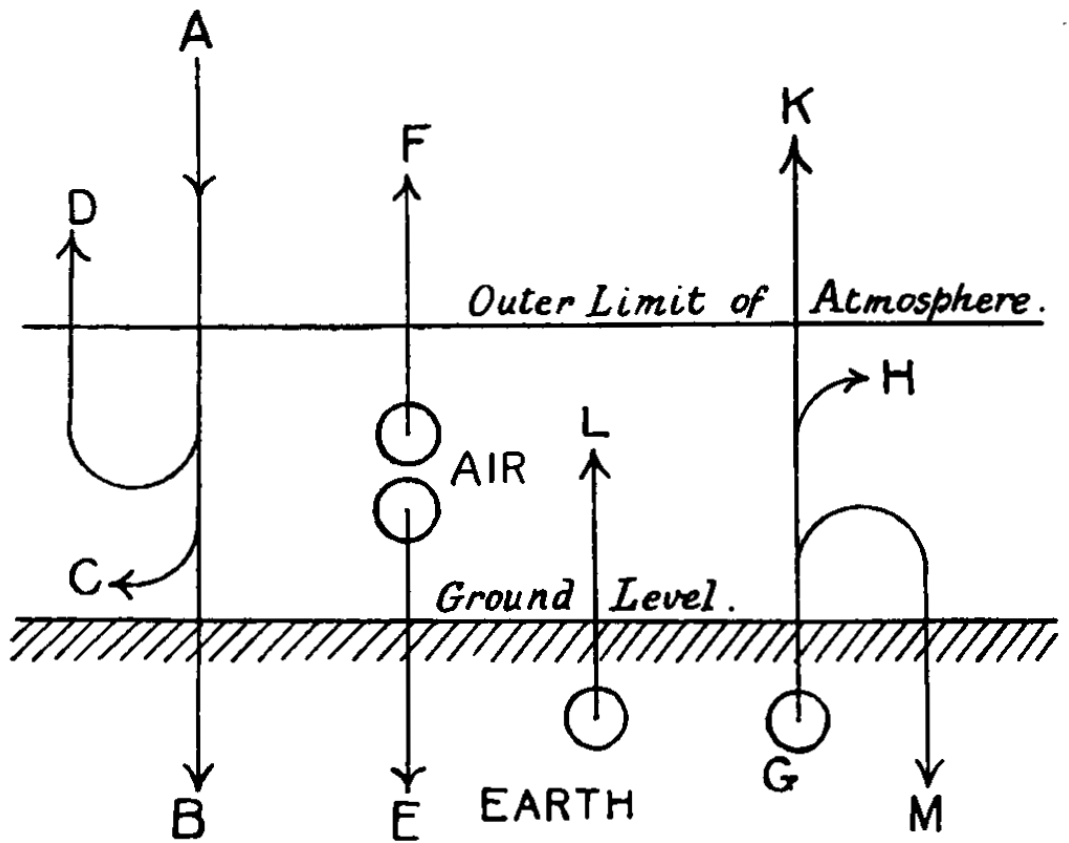

**Figure 4:** Dines' sketch of the radiation balance of the atmosphere.[4] The author described his sketch as follows: Radiant energy, denoted by A, reaches the atmosphere; of this a part (D) is reflected unchanged by the earth or air, a part (C) is absorbed by the air, and a part (B) is absorbed by the Earth. Meanwhile the Earth is radiating its heat (G) outwards; of this let M be reflected back, let H be absorbed and K transmitted. The air is also radiating downwards and upwards; let us call the amounts E and F, and if any part of E is reflected away by the earth it may be included in F. Also heat may pass from the earth to the air or in the opposite direction otherwise than by radiation; let us call this L, earth to air being the positive direction. According to Möller,[30] the quantity M is based on Dines' mistake.

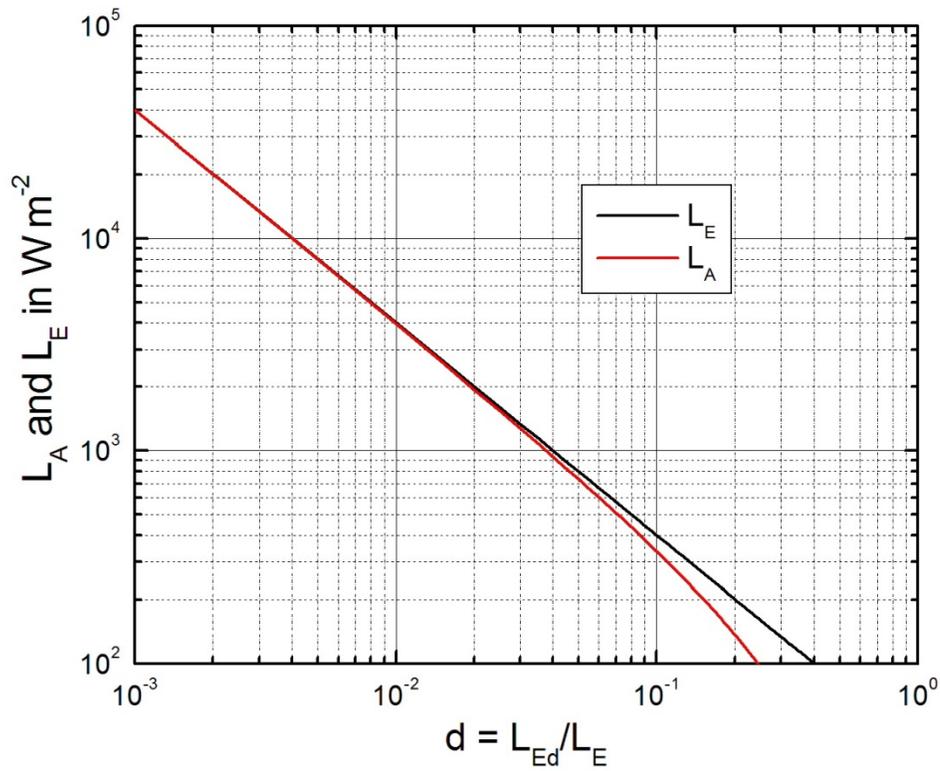

**Figure 5:** Infrared radiative fluxes $L_A$ and $L_E$ emitted by the atmosphere and the Earth's surface, respectively. These fluxes were determined on the basis of Eqs. (17) and (18) of LL here listed as Eqs. (4.8) and (4.9). The value $d = 0.101$ characterizes the flux values $L_A = 333 \, W \, m^{-2}$ and $L_E = 396 \, W \, m^{-2}$ published by Trenberth et al.[12] (see Figure 1). The difference $\Delta L = L_E - L_A$ corresponds to $\Delta L \approx 64 \, W \, m^{-2}$ which is independent of the integral transmissivity $d$. Note that our Figure 5 corresponds to Figure 2 of LL.

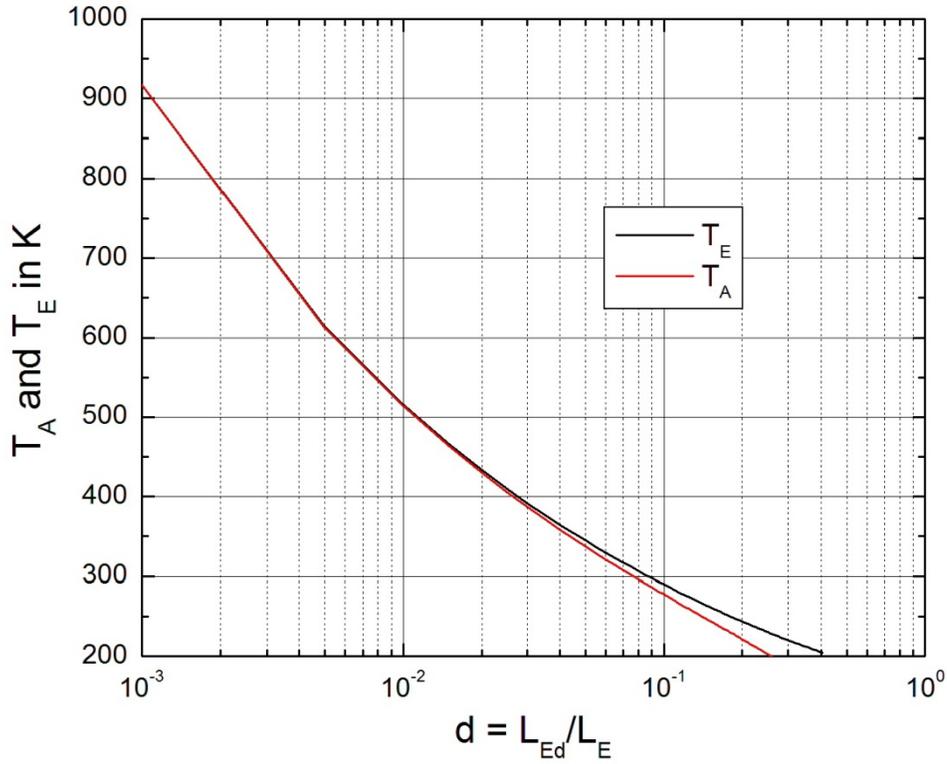

**Figure 6:** As in Figure 5, but for the corresponding temperatures $T_A$ and $T_E$. The value $d = 0.101$ characterizes the temperatures of $T_A = 277$ K and $T_E = 289$ K estimated by LL on the basis of their Eqs. (19) and (20) and the flux values $L_A = 333$ W m$^{-2}$ and $L_E = 396$ W m$^{-2}$ published by Trenberth et al.[12] Note that LL used $\varepsilon_A = 1$ and $\varepsilon_E = 1$, where the former disagrees with the assumption of $d = 0.101$ (see Eq. (4.4)).

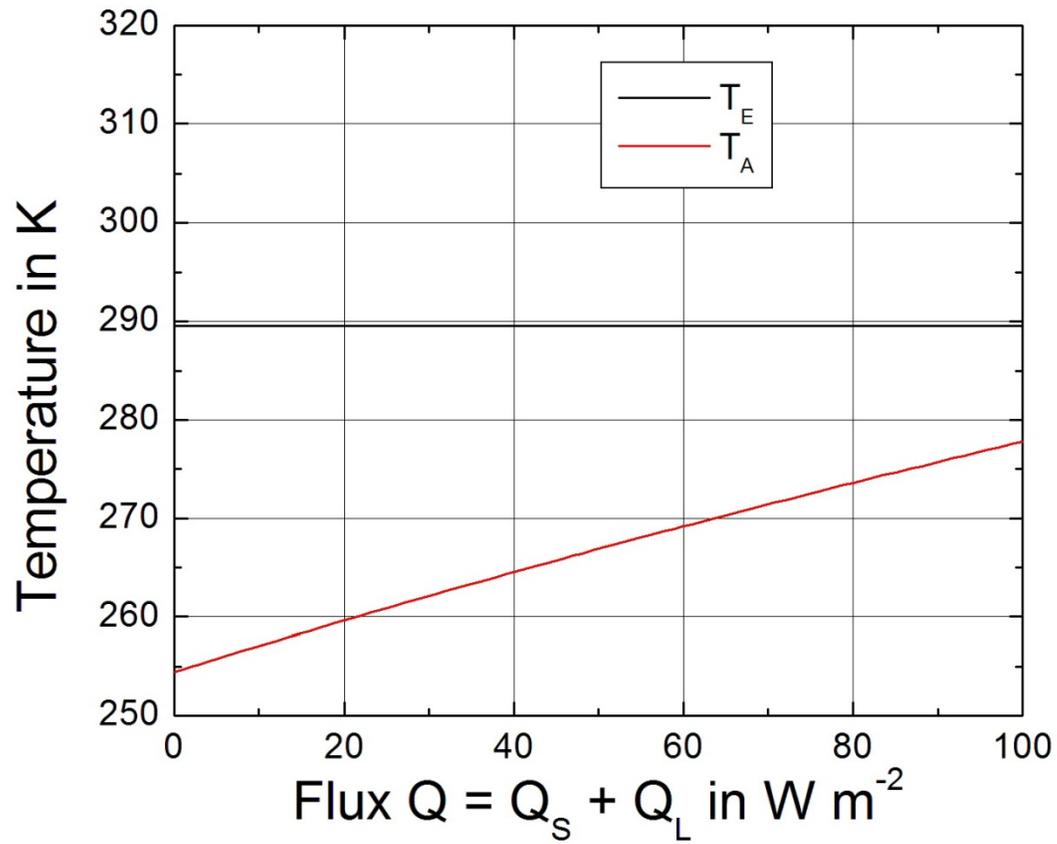

**Figure 7:** Temperatures $T_A$ and $T_E$ as a function of the total heat flux $Q = Q_S + Q_L$. The results are based on Eqs. (17) to (20) of LL.